\documentclass[a4paper,12pt,english]{article}
\pdfoutput=1

\usepackage[pdftex]{graphicx}
\usepackage{graphicx}
\usepackage{colordvi}
\usepackage{fancybox}
\usepackage{cancel}
\usepackage{amsmath}
\usepackage{amssymb}
\usepackage{caption}
\usepackage{appendix}
\usepackage{hyperref} %naredi dokument aktiven (v DVI in PDF)
\usepackage{multirow}
\hypersetup{colorlinks=true, linkcolor=black, urlcolor=blue}
\usepackage{float}
\usepackage{cite}

\usepackage{wasysym}

\usepackage{color}
\definecolor{gray}{rgb}{0.5,0.5,0.5}

\newcommand{\auv}{a_{\mbox{$\scriptscriptstyle UV$}}}
\newcommand{\air}{a_{\mbox{$\scriptscriptstyle IR$}}}

\newcommand{\dir}{\Delta_{\mbox{$\scriptscriptstyle IR$}}}
\newcommand{\aks}{a_{\mbox{$\scriptscriptstyle KS$}}}
\newcommand{\ak}{a_{\mbox{$\scriptscriptstyle K$}}}
\newcommand\lsim{\mathrel{\rlap{\lower4pt\hbox{\hskip1pt$\sim$}}
    \raise1pt\hbox{$<$}}}
\newcommand\gsim{\mathrel{\rlap{\lower4pt\hbox{\hskip1pt$\sim$}}
    \raise1pt\hbox{$>$}}}

\newcommand{\beq}{\begin{equation}}
\newcommand{\eeq}{\end{equation}}
\newcommand{\bea}{\begin{eqnarray}}
\newcommand{\eea}{\end{eqnarray}}
\newcommand{\bem}{\begin{pmatrix}}
\newcommand{\eem}{\end{pmatrix}}
\newcommand{\noi}{\noindent}
\newcommand{\non}{\nonumber}

\newcommand{\bet}{\begin{itemize}}
\newcommand{\eet}{\end{itemize}}
\newcommand{\ben}{\begin{enumerate}}
\newcommand{\een}{\end{enumerate}}

\begin{document}

\numberwithin{equation}{section}

\begin{flushright}
IPPP/18/35 \\
CP3-Origins-2018-019 DNRF90
\end{flushright}

\bigskip

\begin{center}

{\Large\bf Dual RG flows in 4D}
\vspace{1cm}

\centerline{Steven Abel$^{a,}$\footnote{s.a.abel@durham.ac.uk}, Borut Bajc$^{b,}$\footnote{borut.bajc@ijs.si} and 
Francesco Sannino$^{c,}$\footnote{sannino@cp3.sdu.dk}}

\vspace{0.5cm}
\centerline{$^{a}$ {\it\small IPPP, Durham University, South Road, Durham, DH1 3LE}}
\centerline{$^{b}$ {\it\small J.\ Stefan Institute, 1000 Ljubljana, Slovenia}}
\centerline{$^{b}$ {\it\small CP$^3$-Origins \& the Danish IAS, University of Southern Denmark,  Denmark}}
%\centerline{{\it\small }}
 \end{center}

\bigskip

\begin{abstract}
\noindent We present a prescription for using the $a$ central charge to determine the flow of 
a strongly coupled supersymmetric theory from its weakly coupled dual. The approach is based on the equivalence of the scale-dependent $a$-parameter derived from the 
four-dilaton amplitude with the $a$-parameter determined from the Lagrange multiplier method with scale-dependent $R$-charges.
We explicitly demonstrate this equivalence for massive free ${\cal N}=1$ superfields and for weakly coupled SQCD. 
  \end{abstract}

\clearpage

\tableofcontents
% \newpage

\section{Introduction}

Renormalization Group (RG) flow of Quantum Field Theories (QFTs) is thought to be irreversible. In two dimensions this irreversibility is encompassed by the Zamalodchikov $c$-theorem, which states that one can define a monotonically decreasing parameter that interpolates between the central charges $c$ \cite{Zam} of two conformal theories related by an RG flow. An equivalent parameter in 4 dimensions is  Cardy's proposal of the $a$ anomaly, the  coefficient of the Euler density in the trace of the energy momentum tensor \cite{Cardy}. 

In a remarkable paper \cite{Komargodski:2011vj}, Komargodski and Schwimmer (KS) produced a general form for this coefficient to  show that its value will inevitably decrease if a system goes from a UV to an IR fixed point.  The method that was used in \cite{Komargodski:2011vj} is a cousin of 't Hooft anomaly matching, in the sense that a spectator dilaton field is introduced that compensates the anomaly and restores exact Weyl symmetry at all scales, which is spontaneously broken by a dilaton VEV. Using such a set-up the  $a$ parameter can be deduced from the 4 dilaton amplitude. The change in the $a$ parameter between fixed points,  $\air-\auv$, is then found to be always negative by relating it via the optical theorem to the cross-section. Thus the weak form of the $a$-theorem, that  its value will decrease if a system flows from a UV fixed point to an IR one,  can be considered proven. However the strong version, namely that there exists a {\it monotonically} decreasing $a$ function with unambiguous physical meaning all along the flow, appears to be still open  because of the presence of scheme dependent $\beta^2 $ terms in the four dilaton amplitude, as discussed in \cite{rattazzi,Baume:2014rla,Auzzi:2015yia}.

Indeed Jack and Osborn \cite{Osborn:1989td,Jack:1990eb,Osborn:1991gm} showed the existence of a function $\hat{a}$ related to $a$ through the beta functions, that coincides with it at fixed points and that flows with energy scale $\mu$ as   
\begin{equation}
\mu \frac{d \hat{a}}{d\mu } ~=~ \chi_{IJ} \beta^J \beta^I~,
\end{equation} 
where $\beta^I$ are the beta functions of couplings $\lambda_I$, and   $\chi_{IJ}$ is a metric on the space of couplings. 
The problem of proving the monotonicity of the function $\hat{a}$ (and hence the irreversibility of RG flow) is then reduced to one of proving the positive-definiteness of the metric $\chi$ on the space of functions. This problem remains to be solved  (for a review see for example \cite{Shore:2016xor}). 

Our purpose here is to point out that the parametric closeness of $a$ and $\hat{a}$ suggests a method of tracking the approximate flow of a strongly coupled theory. Indeed generally, for flow between nearby fixed points, it seems natural to attempt a perturbative expansion in terms of the beta functions rather than in terms of any couplings \cite{Antipin:2011ny}. In this letter we explore the $a$-parameter as the basis for such an approach, showing how one can use it to follow the flow of arbitrarily strongly coupled SQCD theories between fixed points.

Central to this approach is of course the fact that it is already known how to map the particle content of strongly coupled ``electric'' SQCD theories to weakly coupled ``magnetic'' ones via Seiberg duality \cite{Seiberg:1994pq,Intriligator:1995au}. Thus one can already determine all the discrete parameters of strongly coupled theories, as well as much of their holomorphic data, even when they are away from fixed points. The question we will address here is how one can also determine the flow of the coupling in the strong theory, up to the aforementioned corrections of order $\beta^2$, by mapping from the weak theory. 

The approach continues in the spirit of 't Hooft anomaly matching, by considering the flow of the $a$ parameter. In order to define such a  flow we will use the KS determination of $a$ which involves a certain integration of the 4-dilaton amplitude over the Mandelstam variable  \cite{Komargodski:2011xv}: 
\begin{equation}
\label{eq:ks1}
\auv - \aks(\mu) ~=~ \frac{f^4}{4\pi} \int_{s>\mu^2} \frac{{\rm Im} {\cal A}(s)}{s^3}~ds~,
\end{equation}
where $f$ is the dilaton decay constant and $s$ is the Mandelstam variable, and where we impose an IR cut-off on the integral, $s>\mu^2$, in order to generate a running $a$-parameter, which we denote $\aks(\mu)$. 
The cut-off induced scale dependence in the $a$ parameter interpolates its value smoothly and monotonically between its fixed point values. 
If one supposes that there exist dual descriptions of the entire flow between  the UV and IR fixed points, then the flow induced in the $a$-parameter in the dual theories is identical by the above prescription. 

The route from the $a$ parameters to the couplings is via $R$-charges and hence anomalous dimensions. 
Indeed at the fixed points there already exist well known relations between the anomalous dimensions of fields, their $R$-charges (via the superconformal field theory),  and the  $a$ and $c$ parameters. The latter relations for example take the form\footnote{
%Note that the normalisation in the definition of Eq.\eqref{aparam} is different from that of Eq.\eqref{eq:ks1}; for ease of %comparison and to reduce clutter the expressions will be normalised as they have been in the respective literature.
Here and in the following we will interchangeably use $a$ and $\tilde a$ related by 
\beq
a=\frac{3}{32(4\pi)^2}\,\tilde a\;.
\eeq
}
\begin{equation}
\tilde a\,=\, 
%\frac{3}{32} \left( 
3{\rm Tr} R^3 -{\rm Tr} R
%\right) 
~;~ 
\tilde c\,=\, 
%\frac{1}{32} \left( 
9{\rm Tr} R^3 - 5{\rm Tr} R
%\right)
\, ,
\label{aparam}
\end{equation}
where here $R$ denotes the charges of states contributing to the 't Hooft anomalies (i.e. it would be $R-1$ if the superfield has charge $R$). 

Thus one prescription for defining a set of $R$-charges along the flow is to continue to solve \eqref{aparam} for $R(\mu)$ away from the fixed points, using $\aks(\mu)$ as defined in \eqref{eq:ks1}. We should stress that such a prescription (and the anomalous dimensions it gives rise to) corresponds to a choice of renormalisation scheme.  However as  the right-hand-side of  \eqref{eq:ks1} is the integral of a  physical quantity (namely, by the optical theorem, the 4-dilaton cross-section)  this particular choice has a physical meaning which is similar to that of the ``sliding scale'' scheme \cite{scrucca,Abel:2013mya}. Moreover it is independent of perturbation theory, so it has the same interpretation irrespective of whether one is using the electric or magnetic formulation.  

A second reason to favour ``flowing'' $R$-charges defined in such a way is that they 
appear to coincide with those of the Lagrange-multiplier method suggested by Kutasov \cite{Kutasov:2003ux,Kutasov:2004xu}
\footnote{This method relies on there being a Lagrangian description of the theory, which will be assumed in the following.}. 
The starting point of our discussion will be to demonstrate this unexpected 
equivalence, for flows near fixed points in the Banks-Zaks limit. 
This gives some  physical meaning to the Lagrange-multiplier method when the theory is strongly coupled. Remarkably the RG-scheme implicit in applying \eqref{aparam} to \eqref{eq:ks1},  appears to correspond to that implicit in the Lagrange-multiplier technique\footnote{If along the flow a gauge invariant operator becomes free, a new accidental symmetry arises and one should properly define $\ak(\mu)$ along the lines of \cite{Kutasov:2003iy}.}.  

Consequently one can determine the  $R$-charges of the strongly coupled theory from those 
of the weakly coupled theory, by way of the matched $a$-parameters, which have a well-defined physical meaning 
in terms of the 4-dilaton amplitude, independent of whether the description is strongly or weakly coupled.
From there it is straightforward to determine the anomalous dimensions, hence the NSVZ beta function, and ultimately the gauge coupling in the strongly coupled description.  

\section{Dilaton scattering $a$ versus Lagrange multiplier $a$}

\label{sec:ava}

Let us begin by showing (in the Banks-Zaks limit ) that the $a(\mu)$ parameter one extracts for SQCD 
at scale $\mu$ along the flow between two fixed points using the KS definition \cite{Komargodski:2011xv},  coincides with the Lagrange multiplier $a$-parameter of \cite{Kutasov:2003ux}. 
\begin{figure}[h] 
   \centering
   \includegraphics[width=3.in]{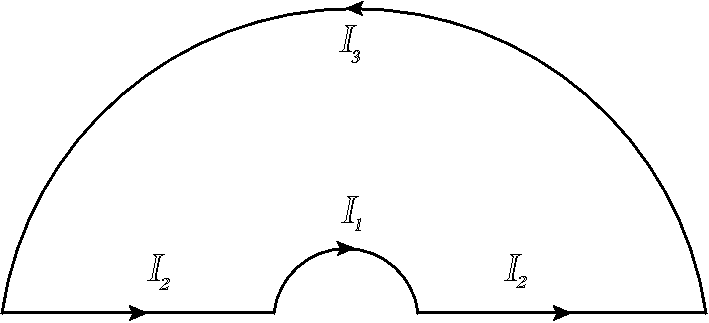} 
    \caption{\it Contour for $\aks(\mu)$.}
   \label{contour}
\end{figure}

First consider  $\aks(\mu)$ in more detail. The prescription of 
\eqref{eq:ks1} can be  understood  
in terms of the contour integral of ${\cal A}/s^3$ around the loop shown in figure \ref{contour}, where the radius of the inner 
contour is $\mu^2$.  The amplitude in this integral is treated as holomorphic in the upper half-plane of complex $s$, with branch-cuts arranged along the real axis. The integral in \eqref{eq:ks1}  corresponds to going along the $I_2$ portion of the contour above the branch-cuts of the amplitude which run along the real axis to plus infinity in the $s$ channel (and minus infinity in the $u$-channel). In the IR the amplitude behaves as 
\begin{equation}
{\cal A}(s) ~=~ 8(\auv-\air) \frac{s^2}{f^4} + {\cal O}
\left( 
\frac{  
m^{2    (4-\dir)   } 
s^{\dir-2} 
}
{f^4}  \right) ,
\label{as1}
\end{equation}
where $\dir>4 $ is the lowest dimension of the irrelevant operators (of the dilaton) in the IR theory, and hence $m$ is the scale of the relevant operators that we added into the UV theory that generated them upon integrating out degrees of freedom.  In the limit that $\mu\rightarrow 0$ we may simply neglect the terms with inverse powers of $m$ (along with $I_3$ which tends to zero) and performing the integral find  by Cauchy's theorem \cite{rattazzi}, 
\begin{equation}
- I_1 ~= ~ \frac{8\pi (\auv-\air)}{f^4} ~=~ I_2 ~=~  2\int_{0}^\infty ds \frac{{\rm Im}{\cal A}(s)}{s^3}\,~ , 
\end{equation}
where we also require Schwartz reflectivity of the Amplitude (namely $\overline{ {\cal A}(s)} =  {\cal A}(\bar{s})$).
This (by way of the optical theorem) is enough to establish the weak $a$-theorem.

By contrast at finite $\mu$ the answer for $I_1$ is of course $\mu $ dependent.
To demonstrate what happens let us first revisit the simple example of free scalar fields of mass $m$ discussed in \cite{Komargodski:2011vj}. Using standard perturbation theory (with the conventions of \cite{Komargodski:2011vj}) their contribution to the 4 dilaton amplitude is found to be 
\beq
\label{eq:abs1}
{\cal A} = -\, 240\, (\auv-\air) ~ \frac{m^4}{f^4} \int _0^1 dx \left( \log(m^2 - s x(1-x) )+ \log(m^2 + s x(1-x) ) \right) + {\rm const.} ~ ,
\eeq
where we assume only these fields contribute to $\auv-\air$. 
The constant term is independent of $s$ and contains counter-terms to remove infinities, but it is not important for the discussion.  Expanding the logarithms in $s/m^2$ and performing the $x$ integral gives the leading contribution in \eqref{as1} (which can be used to check the pre-factor). Alternatively we note that the new absorptive contribution to ${\cal A }$ comes from the region of the integral where the argument of the first logarithm is negative, $s x(1-x) >m^2$. 
Taking $s \rightarrow s + i\epsilon $ in order to be above the branch-cuts,  we find 
\beq
{\rm Im} {\cal A} ~=~ 240 \pi (\auv-\air) \, \frac{m^4}{f^4 }\sqrt{1-4m^2/s} ~. 
\eeq   
Inserting this into the integral $I_2$ with a cut-off then gives a running $a$ parameter  
\begin{eqnarray}
\label{eq:ks2}
\auv - \aks(\mu) ~&=&~ \frac{f^4}{4\pi} \int_{s>\mu^2} \frac{{\rm Im} {\cal A}(s)}{s^3}~ds,\nonumber \\
~&=&~ {(\auv-\air)} \,
(1-\rho {\scriptstyle ( \mu/2m)} )~,
\end{eqnarray}
where 
\beq
\rho(x)~=~  
\begin{cases}
\left( 1-x^{-2} \right)^{3/2}  \left( 1+\frac{3}{2} x^{-2} \right)    ~& ;~~ x  \geq 1 \\
0~  & ;~~ x \leq 1~.
\end{cases}
\eeq
For later  comparison it is useful to rearrange the expression as 
\beq
\label{eq:aks22}
\aks ~=~ \air  ~+~(\auv-\air) \, \rho  {\scriptstyle (\mu/2m)} ~,
\eeq
making it clear that $\rho {\scriptstyle (\mu/2m)}$ correctly scales  the contributions of the scalars to the $\aks$ parameter continuously and monotonically, with $\rho=0$ at $\mu=2m$ to $\rho=1$ at $\mu\rightarrow \infty$.
Thus we may interpret the KS integral of \eqref{eq:ks1} as simply counting the imaginary (absorptive) contributions to the amplitude from states that are able to go on shell when $s>\mu^2$ (a useful reference in this context is  \cite{Kniehl:1996rh}). 

In order to compare the running $\aks$ derived above with the continuously varying $\ak$ function devised for SUSY theories in \cite{Kutasov:2003ux}, we need to extend the simple case above to ${\cal N}=1$ SUSY. Consider the free field theory, consisting of $N_f$ pairs of superfields $\Phi^a $ and $\tilde{\Phi}^a$, $a=1\ldots N_f$. The Lagrangian of \cite{Komargodski:2011vj} can be made supersymmetric in the obvious way, by coupling the fields in a superpotential mass-term $W~\supset ~m \Omega \tilde{\Phi} {\mathbb I}_{\Delta_f\times \Delta_f} \Phi$, where 
$\Omega f $ is the canonically normalised dilaton superfield with $\langle \Omega \rangle=1$. This gives a supersymmetry preserving mass $m$ to $\Delta_f$ pairs of superfields. (As the superpartner of the dilaton does not appear in any loops of interest we can ignore it.) 

The amplitude is of course augmented by superpartner diagrams, but now supersymmetry guarantees that  the coefficient of terms such as  \eqref{eq:abs1} vanish, because otherwise (as these terms  are not zero in the limit of vanishing external momenta and finite  $f$)  they would signal a renormalisation of the superpotential. The non-vanishing terms of interest are, in the standard Passarino-Veltman notation, of the form $s{\cal B}_0 (s,m^2,m^2) $ and friends. Thus the contributions of interest are of the form   
\beq
\label{eq:abs2}
{\cal A} ~=~ - 24\, (\auv-\air) ~ \frac{m^2 s}{f^4} \int _0^1 dx ~\left( \log(m^2 - s x(1-x) )- \log(m^2 + s x(1-x) ) \right) + {\rm const.} ~ ,
\eeq
where as before the second term is really the $u$ Mandelstam variable with $t\rightarrow 0$.

Following the above treatment of the massive scalar, we deduce a running $\aks$ from the absorptive part  which is
\beq
{\rm Im} {\cal A} ~=~  24 \pi (\auv-\air) \, \frac{m^2}{f^4 }~s\sqrt{1-4m^2/s} ~. 
\eeq   
Inserting this  into 
 \eqref{eq:ks1} then gives \eqref{eq:aks22}, but with a modified scaling function,
\beq
\label{eq:aks23}
\rho(x)~=~  
\begin{cases}
\left( 1-x^{-2} \right)^{3/2}      ~& ;~~ x  \geq 1 \\
0~  & ;~~ x \leq 1~.
\end{cases}
\eeq

Let us now compare this expression with the $\ak$ function of \cite{Kutasov:2003ux}. 
The Lagrange multiplier method for this simple case goes as follows.
In general the running $a$-parameter is defined by adding a Lagrange
multiplier for each relevant operator. In this case there is only
one of them, which imposes the constraint from the mass  term. The $a$-function
is therefore given simply by 
\begin{equation}
\tilde a_K ~=~ (N_{f}-\Delta_{f})\left[3(R-1)^{3}-(R-1)\right]+\Delta_{f}\left[3(r-1)^{3}-(r-1)\right]+\lambda\Delta_{f}\left[r-1\right],
\end{equation}
where $R$ is the $R-$charge of the $N_{f}-\Delta_{f}$ chiral superfields
that remain massless and $r$ is the $R$-charge of the last $\Delta_{f}$
flavours, which is considered to be a function of the energy scale. Thus the 
$R$-symmetry we are following along the flow is a linear combination of the superconformal $R$-symmetry 
of the deep $UV$ and the SU$(N_f)\times $SU$(N_f)$ flavour symmetry with which it mixes because of the mass-term (specifically the {\small $diag\left( \,\Delta_f \mathbb{I}_{\scriptscriptstyle N_f-\Delta_f},~(\Delta_f-N_f)  \mathbb{I}_{\scriptscriptstyle \Delta_f} \right)$ } component)\footnote{This is true for $\Delta_f<N_f$: when  
$\Delta_f=N_f$ there is of course no relevant $R$-symmetry left.}.

One first solves to maximise the $a$-function with respect to unfixed $R$-charges, $\frac{\partial a}{\partial r}~=~\frac{\partial a}{\partial R}=0$.
In the absence of the mass-term constraint this simply chooses the free-field
value of $2/3$ for both $R$ and $r$. However at arbitrary Lagrange multiplier
values one finds 
\begin{align}
R & ~=~2/3\,\,,\nonumber \\
r & ~=~1-\frac{\sqrt{1-\lambda}}{3}\,\,.\label{eq:rs-1}
\end{align}
The case where $\lambda=0$ corresponds to $R=r=2/3$ in the deep UV, while $\lambda=1$ 
corresponds to $r=1$, which
is the value forced upon it by the mass-term in the deep IR. Substituting these values into
$\tilde a_K$ we have $\tilde a_{UV}=\frac{2}{9}N_{f}$ and $\tilde a_{IR}=\frac{2}{9}(N_{f}-\Delta_{f})$,
and a running $a-$parameter given by 
\begin{equation}
\tilde a_K~=~\tilde a_{IR}~+~\left(\tilde a_{UV}-\tilde a_{IR}\right)\left(1-\lambda\right)^{\frac{3}{2}}~.
\end{equation}
Comparison with \eqref{eq:aks23} shows that the two $a$-functions precisely coincide if one makes the identification $\lambda ~\equiv ~ \frac{4m^2}{\mu^2}$.
Note that the $a$-functions in the supersymmetric case match essentially because of the non-renormalisation theorem, and that as usual the Lagrange multiplier is essentially the ``coupling'' that 
induces the flow. 

For the SUSY gauge theories of interest the situation is more complicated but the interpretation is always  the same; namely $\aks$ counts the physical states that are able to contribute to the absorptive part of the 4-dilaton amplitude. Meanwhile $\ak$ tracks the mixing of the UV $R$-symmetry with flavour symmetry along the flow \cite{Kutasov:2004xu}. We will now show that at weak coupling, close to the Caswell-Bank-Zaks fixed point,  they are equivalent in this case as well\footnote{It would be interesting to look for reasons behind this equivalence that are valid beyond weak coupling, along the lines of \cite{anselmi1,anselmi2}. For the present work the equivalence in the weakly coupled theories is sufficient.}.

Consider SQCD with $N_f$ flavours of quarks $Q$ and $\tilde{Q}$ flowing from the asymptotically free theory to the fixed point. The $\aks$ parameter was derived in terms of the gauge coupling in \cite{Komargodski:2011xv};
\beq
\aks(\mu)~=~\auv-\frac{N_c^2}{128\pi^2}\int_{g^{-2}(\mu)}^\infty\frac{d\lambda}{\lambda^2}\beta_\lambda\, ,
\eeq
where $\lambda=1/g^2$. 
In the limit $\mu\to0$ this expression reduces to eq. (3.12) of \cite{Komargodski:2011xv}. 
%\beq
%\frac{d\lambda}{\lambda^2}=-dg^2\;\;\;,\;\;\;\beta_{1/g^2}=-b_2(g^2-g_*^2)
%\eeq
%\noi
Using
\beq
b_2 ~=~\frac{N_cN_f}{(8\pi^2)^2}\;\;\;,\;\;\;g_*^2~=~\frac{8\pi^2}{N_f}\epsilon
%\;\;\;,\;\;\;\epsilon~=~\frac{3N_c-N_f}{N_c}~,
~,
\eeq
where 
\beq
\epsilon~=~\frac{3N_c-N_f}{N_c}~\ll ~1~,
\eeq
the integral gives
\beq
\label{akomar}
\aks(\mu)~=~\auv\,-\,\frac{N_c^3N_f}{32(8\pi^2)^3}g^2(\mu)(2g_*^2-g^2(\mu))~,
\eeq
with $g^2(\mu)$ being a solution of the 2-loop RGE,
\beq
\frac{dg^2}{d\log{\mu}}~=~b_2g^4(g^2-g_*^2)~.
\eeq

Again we can compare this parameter to the continuously varying $\ak$-function  of \cite{Kutasov:2003ux}. 
In an $SU(N)$ gauge theory it can  be written in generality as
\beq
\label{lagrange}
\tilde a_K~=~2(N_c^2-1)+\sum_i|r_i|\,\left(a_1(R_i)-(R_i-R_i^{\scriptscriptstyle  IR})\,a_1'(R_i)\right)~,
\eeq
where $|r_i|$ is the dimension of the representation $r_i$, the prime means derivative with respect to $R$ and 
where
\beq
a_1(r)~\equiv~3(r-1)^3-(r-1)\, .
\eeq
In the case of electric SQCD this gives 
\beq
\label{akfromR}
\tilde a_K(\mu)~=~2(N_c^2-1)+2N_cN_f\left(a_1(R_Q)-(R_Q-R_Q^*)a_1^\prime(R_Q)\right)\, .
\eeq
In order to compare with $\aks$ we relate the $R$-charges to the anomalous dimensions through 
\beq
\label{r-gam}
R_Q~=~\frac{2}{3}\left(1+\frac{\gamma_Q}{2}\right)~.
\eeq
This equation holds along the flow, but only at the endpoints of the flow does $R_Q$ coincide with the respective 
super-conformal $R$-charges of the fixed points. 
\noi
The anomalous dimension can be perturbatively calculated at 1-loop as
\beq
\gamma_Q~=~-\frac{N_c}{8\pi^2}g^2~,
\eeq
and then using
\beq
R_Q^{\scriptscriptstyle UV}=\frac{2}{3}\;\;\;,\;\;\;R_Q^*=1-\frac{N_c}{N_f}~,
\eeq
we easily find the same leading contribution as that in  (\ref{akomar}), and hence  
$\ak\equiv \aks$ as we wished to prove.

At this point one could ask, what is the meaning of equating a scheme independent quantity such as $\aks(\mu)$ 
with a scheme dependent one such as $\ak(\mu)$. This is of course what we always do when we calculate 
a cross section (in which we are bound to choose a scheme), and compare it to its (scheme independent) measured 
value. The theoretical result becomes scheme independent only when all terms in perturbation theory 
are taken into account, but never at finite order. Therefore the equivalence is only a perturbative one. Nevertheless as 
we shall now show, what it does do is allow us to develop a perturbative description of the flow in a strongly coupled theory.

\section{A perturbative calculation of a non-perturbative flow}
\label{sec:flow}

We now wish to explore how this equivalence can be used to determine the gauge coupling flow in a strongly coupled description.   
To do so we will consider a strongly coupled SQCD (in the conformal window) when one invokes a flow by adding a mass term for one flavour, and will make use of the well-known duality between 
this theory and Higgsing in a weakly coupled magnetic description, described in \cite{Seiberg:1994pq}\footnote{For a pedagogical 
description of such a set-up see for example \cite{Strassler:2005qs}.}. 

The original electric SQCD theory is an ${\cal N}=1$ SU$(N_c)$ theory with $N_f+1$ flavours of $Q$ and $\tilde{Q}$ quarks and anti-quarks. We add a mass-term of the form 
\beq
 W_e ~=~ m\, Q_{N_f+1}\tilde Q^{N_f+1}
\eeq
in its superpotential. In the IR, i.e. at energy below $m$, it flows to a new theory with $N_f$ flavours, hence effectively there is 
a UV fixed point with $N_f+1$ flavours at energy above $m$, and an IR fixed point with $N_f$ flavours. If we take $2 N_f = {3} N_c +1$ then the theory is expected to be strongly coupled for large $N_c$ all along the flow. 

Meanwhile the magnetic description is an SU$(\tilde{N}_c+1)$ theory with $\tilde{N}_c=N_f-N_c$ and, as well as $N_f+1$  
flavours of quarks $q$ and $\tilde{q}$, it contains an elementary $(N_f+1)\times (N_f+1)$ meson $\Phi$ formed from a composite of the electric quarks,  which we will take to be $\Phi \equiv \frac{1}{     {{\Lambda}}     }   {Q\cdot \tilde{Q}}$ where ${\Lambda}$ is the dynamical scale of the theory, and a superpotential 
\beq
\label{magsup}
 W_m~=~ m\, {\Lambda} \Phi_{N_f+1}^{N_f+1}+\tilde{y} \, \Phi\tilde{q}\cdot q~,
\eeq
whose first term derives from the mass-term, and where the Yukawa coupling is $\tilde{y}={\Lambda}/\hat{\Lambda}$ with   
$\hat{\Lambda}\sim \Lambda$. The magnetic theory which has $N_f=3{\tilde{N}}_c-1$ is arbitrarily weakly coupled, so its flow can be followed perturbatively. In particular the linear meson term in the superpotential causes a Higgsing down to ~SU$({\tilde{N}}_c)$.  
For completeness we summarise the flows as seen in the two dual theories in Table \ref{theories}, where the RG scale is defined with respect to $m$, that is 
\beq
t~\equiv~\log{(\mu/m)}~.
\eeq

We can easily determine the difference 
between the UV and IR $a$-central charges 
\bea
\tilde a_{UV}-\tilde a_{IR} &~=~&2N_c(N_f+1)a_1(1-N_c/(N_f+1))-2N_cN_fa_1(1-N_c/N_f)\non\\
&~=~&\frac{6N_c^2(2N_f+1)}{N_f^2(N_f+1)^2}~,
\eea
which is positive for all $N_f>0$, and thus the weak $a$-theorem is satisfied.\\

%%%%%%%%%%%%%%%%%%%%%%%%%%%%%%%%%%%%%%%%%%%%%%%%%%%%%%%%%%%%%%%%%
\begin{center}
\begin{tabular}{| c || c | c |} 
 \hline
& IR\,($t<0$) & UV\,($t>0$) \\ [0.5ex] 
 \hline\hline
\multirow{2}{4.5em}{magnetic theory} 
& $N_f$ flavours & $N_f+1$ flavours \\ 
& $\tilde N_c$ colours & $\tilde N_c+1$ colours \\
\hline
\multirow{2}{4.5em}{electric theory} 
& $N_f$ flavours & $N_f+1$ flavours \\ 
& $N_c$ colours & $N_c$ colours \\
\hline
  \hline
\end{tabular}
\captionof{table}
{\label{theories} \it The dual theories considered in the text with $N_c=N_f-\tilde N_c$. We consider throughout the case of  $N_f~=~\frac{3}{2}N_c + \frac{1}{2} ~=~ 3\tilde N_c-1$.}
\end{center}
%%%%%%%%%%%%%%%%%%%%%%%%%%%%%%%%%%%%%%%%%%%%%%%%%%%%%%%%%%%%%%%%%

As discussed, our aim is to determine the gauge coupling for the original strongly coupled electric theory.
In order to do this we first consider in detail the dual of the UV theory, and the dual of the IR theory, both of which are known. By choosing $N_f\approx 3\tilde N_c$  
and large $\tilde N_c\equiv N_f-N_c$, the magnetic theory\footnote{It is convenient to take the magnetic 
theory to be perturbative as then only one 
coupling - the electric gauge coupling - is non-perturbative and thus the matching of $a$-parameters determines it uniquely.} is made perturbative both in the UV and the IR so we can calculate its flow 
with good accuracy along the whole RG trajectory. As we also know the (in principle non-perturbative) 
interacting electric theories in both the UV and the IR, we assume that the flow of the magnetic theory is dual to that of the strongly coupled electric theory along the whole trajectory.

Before entering into the explicit computation, let us clarify the idea and the procedure we will follow. The magnetic theory is perturbative and 
is thus under control in the whole region 
between the deep UV ($\mu=+\infty$, $t=+\infty$) and the deep IR ($\mu=0$, $t=-\infty$). The gauge and Yukawa couplings are 
continuous along the whole flow, while the two beta functions and the two anomalous dimensions are, due to the mass independent 
character of the NSVZ scheme, discontinuous at the explicit quark mass scale ($\mu=m$, $t=0$). 

However in one half of the flow ($\mu<m$) we are able to use the $a$ parameters to  track (perturbatively and in the NSVZ scheme) 
the evolution of the strongly coupled theory from that in the weakly coupled one. 
In particular the electric theory is non-perturbative and what we  know from Seiberg duality is that it is equivalent to the 
magnetic theory at $\mu=m$ ($t=0$), which becomes the ``new'' UV, and in the deep IR at $\mu=0$ ($t=-\infty$). By continuity\footnote{For a 
discussion on this point see for example \cite{Strassler:2005qs}.} we assume that 
the perturbative magnetic and non-perturbative electric theories are dual between these two endpoints. 
In this region the physical quantity $\aks$ is by definition the same in the magnetic and electric theory, and as motivated 
in the previous section we equate $\ak$ and $\aks$, allowing us to explicitly calculate the 
central charge in the whole energy range $0<\mu<m$ in the magnetic theory, as a perturbative function of 
the gauge and Yukawa coupling constants. This (via the equivalence of the $a$ parameters) yields the non-perturbative $R$-charges and hence anomalous dimensions of the strongly coupled electric theory, which in turn yields 
the explicit numerical solution of the RGE for the electric gauge coupling constant.

\subsection{UV ($0<t<\infty$): magnetic theory}

We start in the UV with the magnetic theory, which is an SU($\tilde N_c+1$) gauge theory with $N_f+1$ 
quarks $ q+\tilde q$ and $(N_f+1)^2$ singlet meson fields with the superpotential of \eqref{magsup}.
We will work in terms of 
\beq
\tilde\alpha_g\equiv\frac{(\tilde N_c+1)\tilde g^2}{(4\pi)^2}\;\;\;,\;\;\;
\tilde\alpha_y\equiv\frac{(\tilde N_c+1)\tilde y^2}{(4\pi)^2}~.
\eeq
%It is useful to introduce a small parameter $\epsilon$ defined such that  
%
%\beq
%N_f+1=\tilde N_c(3-\epsilon)
%\eeq
The theory is asymptotically free when $N_f+1 = 3\tilde{N}_c$ since 
\beq
\label{b1UV}
b_1~=~3(\tilde N_c+1)-(N_f+1)~
%=2+\epsilon\tilde N_c
=~3~,
\eeq
and at $t\to\infty$ all couplings go to zero. For $\mu\gg \Lambda$ the 1-loop approximation is sufficient, and one has 
the usual evolution with dynamical scale given by  $\Lambda \equiv \mu\exp\left( -\frac{1}{2b_1
\tilde{\alpha} _g(\mu)}\right)$.
Towards the IR the flow approaches a Banks-Zaks fixed point that for larger $\tilde{N}_c$ becomes increasingly perturbative. 
Indeed the two-loop RGEs (see for example \cite{Martin:1993zk,Antipin:2011ny})
give the fixed points to be at    
\begin{align}
\tilde\alpha_g(0^+)~=~&
\frac{\left(2\frac{N_f+1}{\tilde N_c+1}+1\right)\left(\frac{N_f+1}{\tilde N_c+1}-3\right)}{6+8\frac{N_f+1}{\tilde N_c+1}-4\left(\frac{N_f+1}{\tilde N_c+1}\right)^2+2\frac{N_f+1}{(\tilde N_c+1)^3}}
~\xrightarrow{N_f\to3\tilde N_c-1}~\frac{7}{2\tilde N_c}+\frac{43}{2\tilde N_c^2}+{\cal O}(1/\tilde N_c^3)~,\nonumber \\
\tilde\alpha_y(0^+)~=~&\frac{2\left(1-\frac{1}{(\tilde N_c+1)^2}\right)\left(\frac{N_f+1}{\tilde N_c+1}-3\right)}{6+8\frac{N_f+1}{\tilde N_c+1}-4\left(\frac{N_f+1}{\tilde N_c+1}\right)^2+2\frac{N_f+1}{(\tilde N_c+1)^3}}
~\xrightarrow{N_f\to3\tilde N_c-1}~\frac{1}{\tilde N_c}+\frac{7}{\tilde N_c^2}+{\cal O}(1/\tilde N_c^3)~.
\end{align}
Defining 
\beq
\Delta^{\scriptscriptstyle (+)}_{\tilde{g}}(t) ~=~ \tilde\alpha_g(t)-\tilde\alpha_g(0^+)~,~~
\Delta^{\scriptscriptstyle (+)}_{\tilde{y}}(t) ~=~ \tilde\alpha_y(t)-\tilde\alpha_y(0^+)~,
\eeq
the 2-loop RGEs can be rephrased as
{\small 
\bea
\frac{d\tilde\alpha_g(t)}{dt}&=&-2\tilde\alpha_g^2(t) \times \left[\left(6-4\frac{N_f+1}{\tilde N_c+1}+2\frac{N_f+1}{(\tilde N_c+1)^3}\right)
\Delta^{\scriptscriptstyle (+)}_{\tilde{g}}(t)
+2\left(\frac{N_f+1}{\tilde N_c+1}\right)^2\Delta^{\scriptscriptstyle (+)}_{\tilde{y}}(t)\right]~,\nonumber \\
\frac{d\tilde\alpha_y(t)}{dt}&=&2\tilde\alpha_y(t)\times  \left[-2\left(1-\frac{1}{(\tilde N_c+1)^2}\right)\Delta^{\scriptscriptstyle (+)}_{\tilde{g}}(t)
+\left(2\frac{N_f+1}{\tilde N_c+1}+1\right)\Delta^{\scriptscriptstyle (+)}_{\tilde{y}}(t)\right]~. 
\label{eq:rges}
\eea
}

\noindent The terms proportional to $\tilde{\alpha}_g(0^+)$ and $\tilde{\alpha}_y(0^+)$ in these expressions are the one-loop terms, while the remaining terms are two-loop.

Finally we can calculate the R-charges at the $t=0^+$ fixed point:
\beq
R_q(0^+)~=~1-\frac{\tilde N_c+1}{N_f+1}\;\;\;,\;\;\;R_\Phi(0^+)~=~2-2R_q(0^+)~,
\eeq

\noi
which are perturbative (free) $R_q=R_\Phi=2/3$ in the large $\tilde N_c$ limit with $N_f\rightarrow 3\tilde N_c-1$, in accord with the magnetic theory being parametrically perturbative for all positive $t$.

\subsection{UV ($0<t<\infty$): electric theory}

Apart from the far UV (as it is also asymptotically free) the form of the electric dual theory is known 
only in the $t\to0^+$ limit, where it 
is an SU($N_c$) gauge theory with 
%$N_c=(N_f+1)-(\tilde N_c+1)=N_f-\tilde N_c$ and with 
$N_f+1$ quarks $Q+\tilde Q$ and vanishing superpotential. In the same limit the 
fixed point determines the value of the R-charge:
\beq
R_Q(0^+)~=~1-\frac{N_c}{N_f+1}~ .
\eeq
In the large $N_c$ limit (as we have $N_f=\frac{3}{2} N_c+1$) 
this value is clearly interacting, $R_Q\rightarrow 1/3$, in accord with the composite 
object $Q\cdot \tilde{Q}$ becoming free. We do not know the value of the electric gauge coupling at other values of $t$.

\subsection{IR ($-\infty<t<0$): magnetic theory}

\begin{figure}[h] 
   \centering
   \includegraphics[width=3.in]{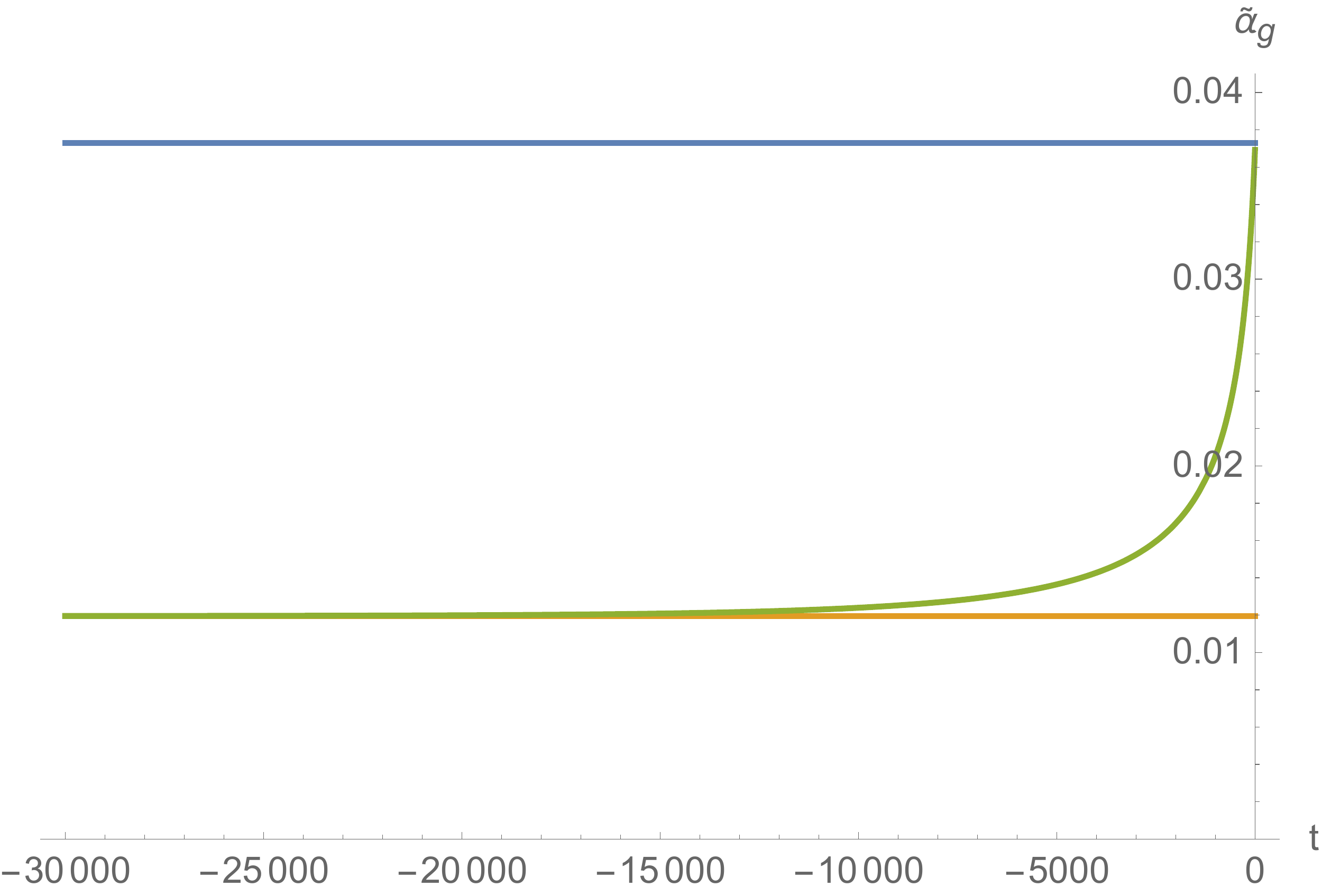} 
   \includegraphics[width=3.in]{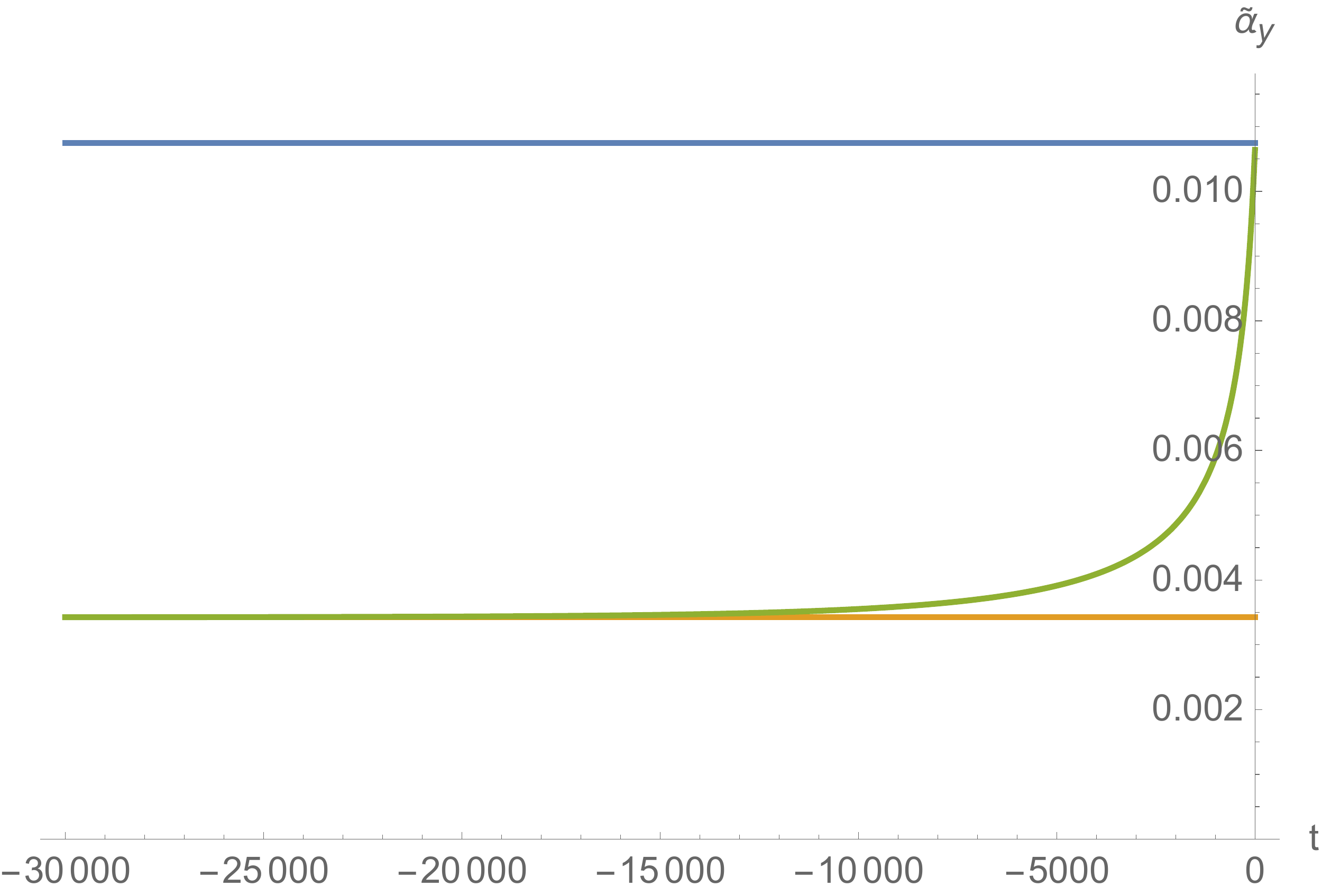} 
   \caption{\it The perturbative running of the gauge (left) and Yukawa (right) coupling constants of the 
   magnetic theory  for $\tilde N_c=100$ and $N_f=3\tilde N_c-1$, from the UV fixed point ($t=0$ with $N_f+1$ quarks and $\tilde N_c+1$ colours), to the IR fixed point ($t=-\infty$ with $N_f$ quarks and $\tilde N_c$ colours). The lower and upper 
   lines denote the UV and IR values of the couplings.}
   \label{gymag}
\end{figure}

We now turn to the flow of interest, towards the IR, for $t<0$. Here the magnetic theory is an SU($\tilde N_c$) gauge 
theory with $N_f$ quarks $q+\tilde q$ and $N_f\times N_f$ gauge singlet mesons, which for convenience we continue to call $\Phi$. At $t=0$ the boundary 
conditions of the couplings,
\beq
\label{defgy}
\tilde\alpha_g~\equiv~\frac{\tilde N_c\tilde g^2}{(4\pi)^2}\;\;\;,\;\;\;
\tilde\alpha_y~\equiv~\frac{\tilde N_c\tilde y^2}{(4\pi)^2}~,
\eeq
are determined by continuity\footnote{Note that we are using the NSVZ scheme which is a mass independent scheme: 
this means that the perturbative gauge couplings are continuous passing the mass scale, while the beta functions or anomalous 
dimensions are not. The apparent discontinuity in (\ref{discat0}) is clear from the way the coupling constants are defined in 
(\ref{defgy}), i.e. with the discontinuity being in the number of colours.},
\bea
\label{discat0}
\tilde\alpha_g(0^-)&=&\frac{\tilde N_c}{\tilde N_c+1}\tilde\alpha_g(0^+)~,\nonumber \\
\tilde\alpha_y(0^-)&=&\frac{\tilde N_c}{\tilde N_c+1}\tilde\alpha_y(0^+)~.
\eea
The flow of the magnetic theory can be determined perturbatively from the RGEs. Defining 
\beq
\Delta^{\scriptscriptstyle (-)}_{\tilde{g}}(t) ~=~ \tilde\alpha_g(t)-\tilde\alpha_g(-\infty)~,~~
\Delta^{\scriptscriptstyle (-)}_{\tilde{y}}(t) ~=~ \tilde\alpha_y(t)-\tilde\alpha_y(-\infty)~,
\eeq
where the new fixed point is at 
\begin{align}
%\label{alphagIR}
\tilde\alpha_g(-\infty)~=~&
\frac{\left(2\frac{N_f}{\tilde N_c}+1\right)\left(\frac{N_f}{\tilde N_c}-3\right)}{6+8\frac{N_f}{\tilde N_c}-
4\left(\frac{N_f}{\tilde N_c}\right)^2+2\frac{N_f}{\tilde N_c^3}}
~\xrightarrow{N_f\to3 \tilde N_c-1}~\frac{7}{6\tilde N_c}+\frac{25}{9\tilde N_c^2}+{\cal O}(1/\tilde N_c^3)  ~, \non \\
\label{alphayIR}
\tilde\alpha_y(-\infty)~=~&
\frac{2\left(1-\frac{1}{\tilde N_c^2}\right)\left(\frac{N_f}{\tilde N_c}-3\right)}{6+8\frac{N_f}{\tilde N_c}-4\left(\frac{N_f}{\tilde N_c}\right)^2+2\frac{N_f}{\tilde N_c^3}}
~\xrightarrow{N_f\to3\tilde N_c-1}~\frac{1}{3\tilde N_c}+\frac{8}{9\tilde N_c^2}+{\cal O}(1/\tilde N_c^3)~,
\end{align}
they are 
{\small{\bea
\label{rgeIRgmag}
\frac{d\tilde\alpha_g(t)}{dt}&=&-2\tilde\alpha_g^2(t) \times \left[\left(6-4\frac{N_f}{\tilde N_c}+2\frac{N_f}{\tilde N_c^3}\right)\Delta^{\scriptscriptstyle (-)}_{\tilde{g}}(t)
+2\left(\frac{N_f}{\tilde N_c}\right)^2\Delta^{\scriptscriptstyle (-)}_{\tilde{y}}(t)\right]~,\non\\
\label{rgeIRymag}
\frac{d\tilde\alpha_y(t)}{dt}&=&2\tilde\alpha_y(t)\times \left[-2\left(1-\frac{1}{\tilde N_c^2}\right)\Delta^{\scriptscriptstyle (-)}_{\tilde{g}}(t)
+\left(2\frac{N_f}{\tilde N_c}+1\right)\Delta^{\scriptscriptstyle (-)}_{\tilde{y}}(t)\right]~.
\eea}}\\
The evolution is 
shown in Fig.~\ref{gymag}
for $\tilde N_c=100$ and $N_f=3\tilde N_c-1$. 

\begin{figure}[h] 
   \centering
   \includegraphics[width=3.in]{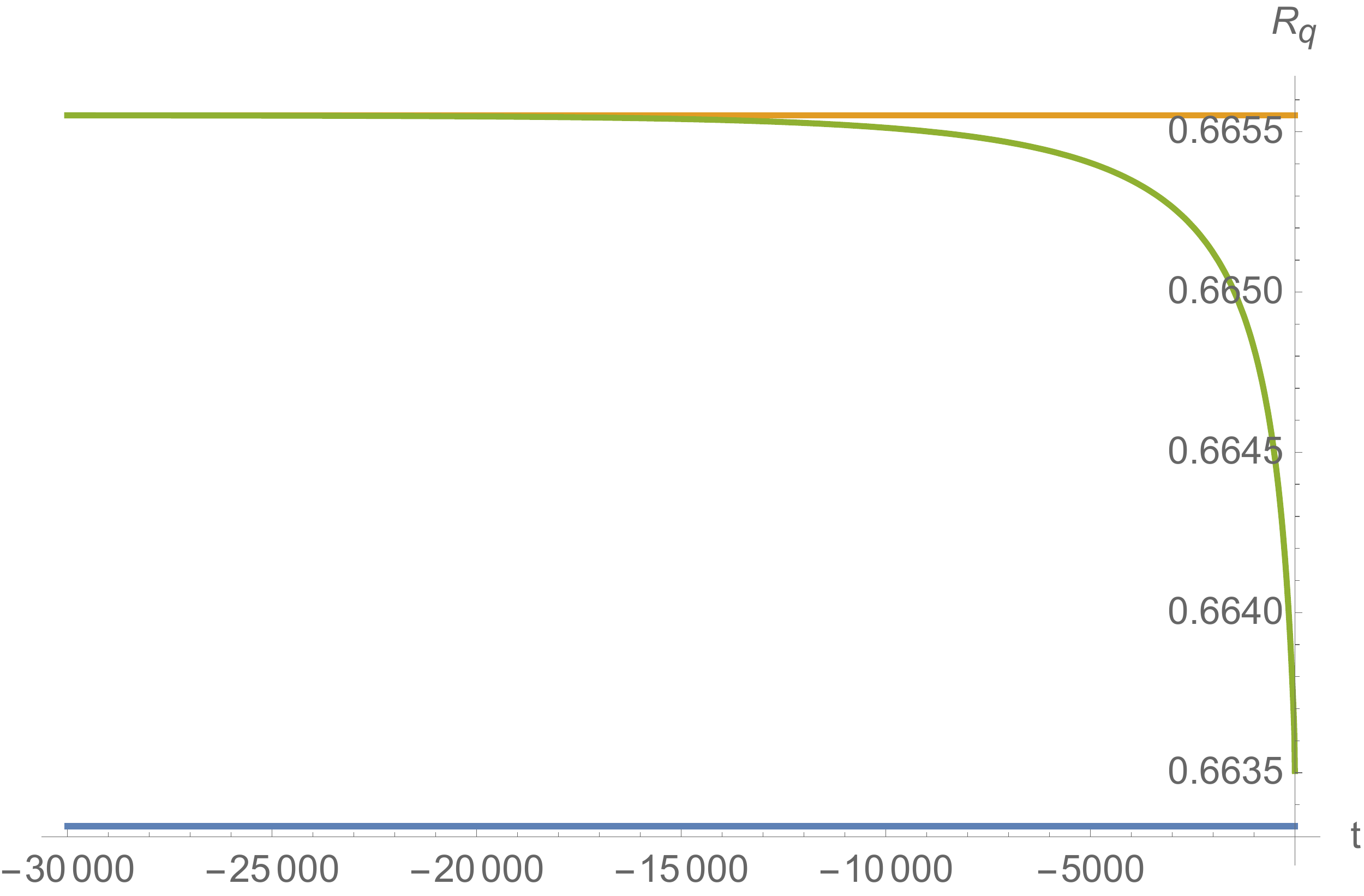} 
   \includegraphics[width=3.in]{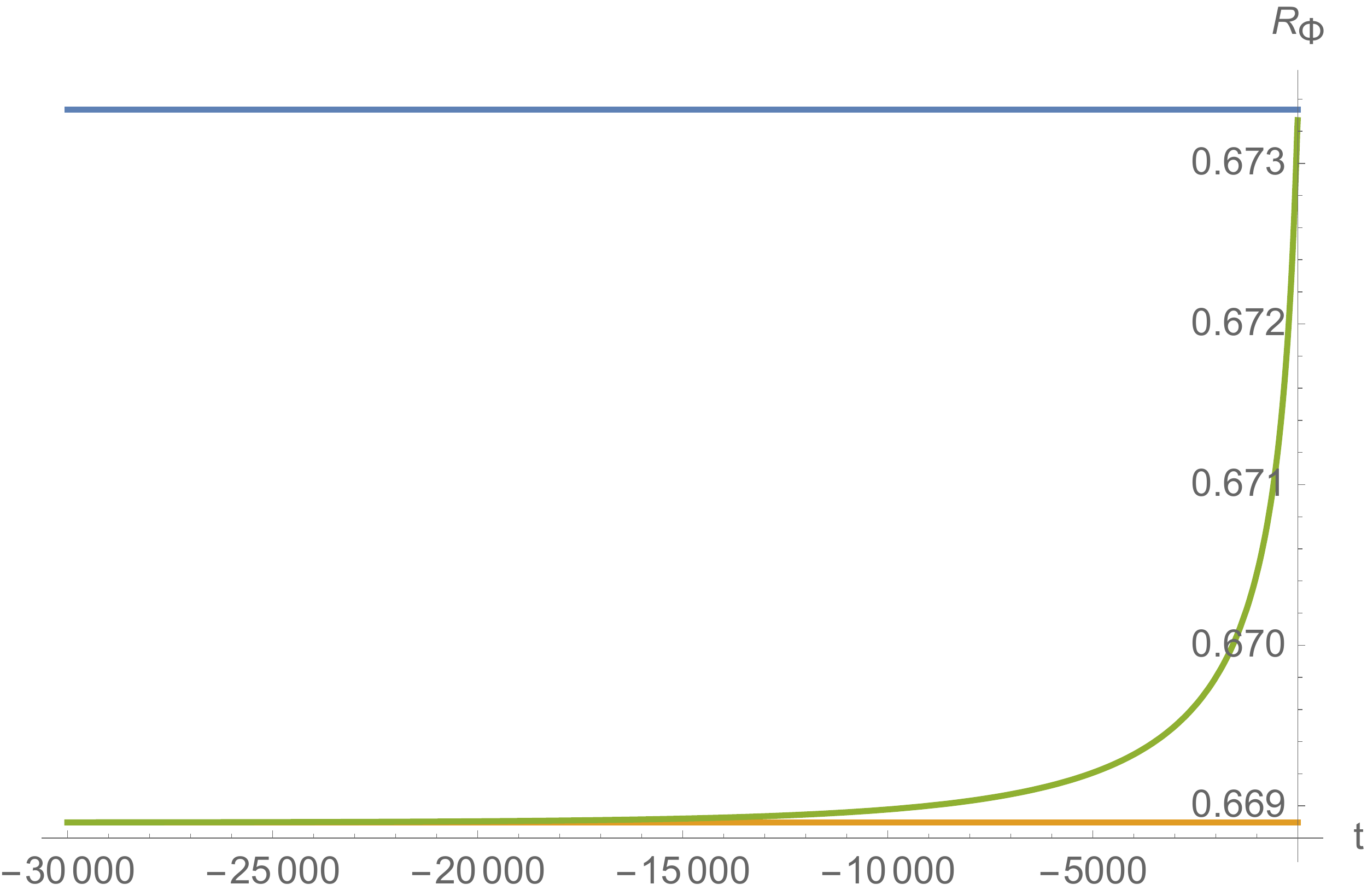} 
   \caption{\it 
  The ``flowing'' R-charges (green) of the quark (left and meson (right) in the magnetic SQCD with gauge SU($\tilde N_c$) 
  and $N_f$ quarks $q+\tilde q$ and $N_f^2$ mesons $\Phi$, with $N_f=3\tilde N_c-1$. The flow has been 
  found using the perturabative relations (\ref{Rq}) and (\ref{RM}) and using $\tilde N_c=100$. The blue straight 
  lines are the values $R_q(0^+)$ and $R_\Phi(0^+)$ obtained in the fixed point above the mass $m$. 
  Notice that the values $R_q(0^-)$ and $R_\Phi(0^-)$ do not coincide with them: although the gauge couplings 
  $\tilde g$ and $ \tilde y$ are continuous, 
  %neither the combinations $\tilde \alpha_g$ and $\tilde\alpha_y$, not the $R$-charges are
  the $R$-charges are not: they are in some sense proportional to the non-continuous beta-functions. Finally the orange 
  straight lines are the limiting values $R_q(-\infty)$ and $R_\Phi(-\infty)$ obtained from the IR fixed point couplings.}
   \label{RqRMmag}
\end{figure}

It is useful to explicitly express the flowing $R$-charges in terms of the couplings. This we can do 
because the theory is perturbative (approximately, order by order in perturbation theory). 
From the usual definition 
of the  NSVZ beta function and the relation in \eqref{r-gam}, we have 
\bea
\label{Rq}
\beta(\tilde\alpha_g)&=&-6\tilde\alpha_g^2f(\tilde\alpha_g)\left(1+\frac{N_f}{\tilde N_c}\left(R_q-1\right)\right)\non ~,\\
\label{RM}
\beta(\tilde\alpha_y)&=&3\tilde\alpha_y\left(2R_q+R_\Phi-2\right)~,
\eea
with
\beq
\label{f}
f(x)\equiv\frac{1}{1-2x}~.
\eeq
Comparison with the r.h.s. of (\ref{rgeIRgmag})  gives
\begin{align}
R_q(t)-R_q(-\infty)~&~=~-\frac{2}{3}\left(1-\frac{1}{\tilde N_c^2}\right)\Delta^{\scriptscriptstyle (-)}_{\tilde{g}}~
+\frac{2}{3}\frac{N_f}{\tilde N_c}\Delta^{\scriptscriptstyle (-)}_{\tilde{y}} 
+~{\cal O} \left( \Delta^2\right) ~, \non \\
~~R_\Phi(t)-R_\Phi(-\infty)&~=~\frac{2}{3}\Delta^{\scriptscriptstyle (-)}_{\tilde{y}} +~{\cal O} \left(\Delta^2\right) ~. 
\label{Rqalpha}
\end{align}Their evolution for $t<0$ is shown in Fig.~\ref{RqRMmag}. Note that although we do not display them explicitly, the order $\Delta^2$ terms as derived from Eq.\eqref{Rq}, are actually required later in order to get consistent convergence to the IR fixed point in the strongly coupled description. 

\subsection{IR ($-\infty<t<0$): electric theory}

Up to this point, for $t<0$, everything has been perturbative. Now let us now consider the original electric theory in the range 
$-\infty<t<0$. In the limit $t\to-\infty$ the theory is SU($N_c$) SQCD with $N_f$ quarks 
$Q+\tilde Q$ and no superpotential. 
%Right above the mass, i.e. at $t\to0^+$ the electric theory is again known: 
%as stated above, it is a SU($N_c$) gauge theory with $N_c=N_f-\tilde N_c$ and $N_f+1$ quarks $Q+\tilde Q$. 

Let us assume that the same pair of dual theories describe the physics along the whole RGE running. 
As the parameter $\aks$ is a function of the amplitude its definition is independent of which description is being used and hence 
its value in the electric and magnetic theories is the {same all along the  
flow}. 

\begin{figure}[h] 
   \centering
   \includegraphics[width=3.in]{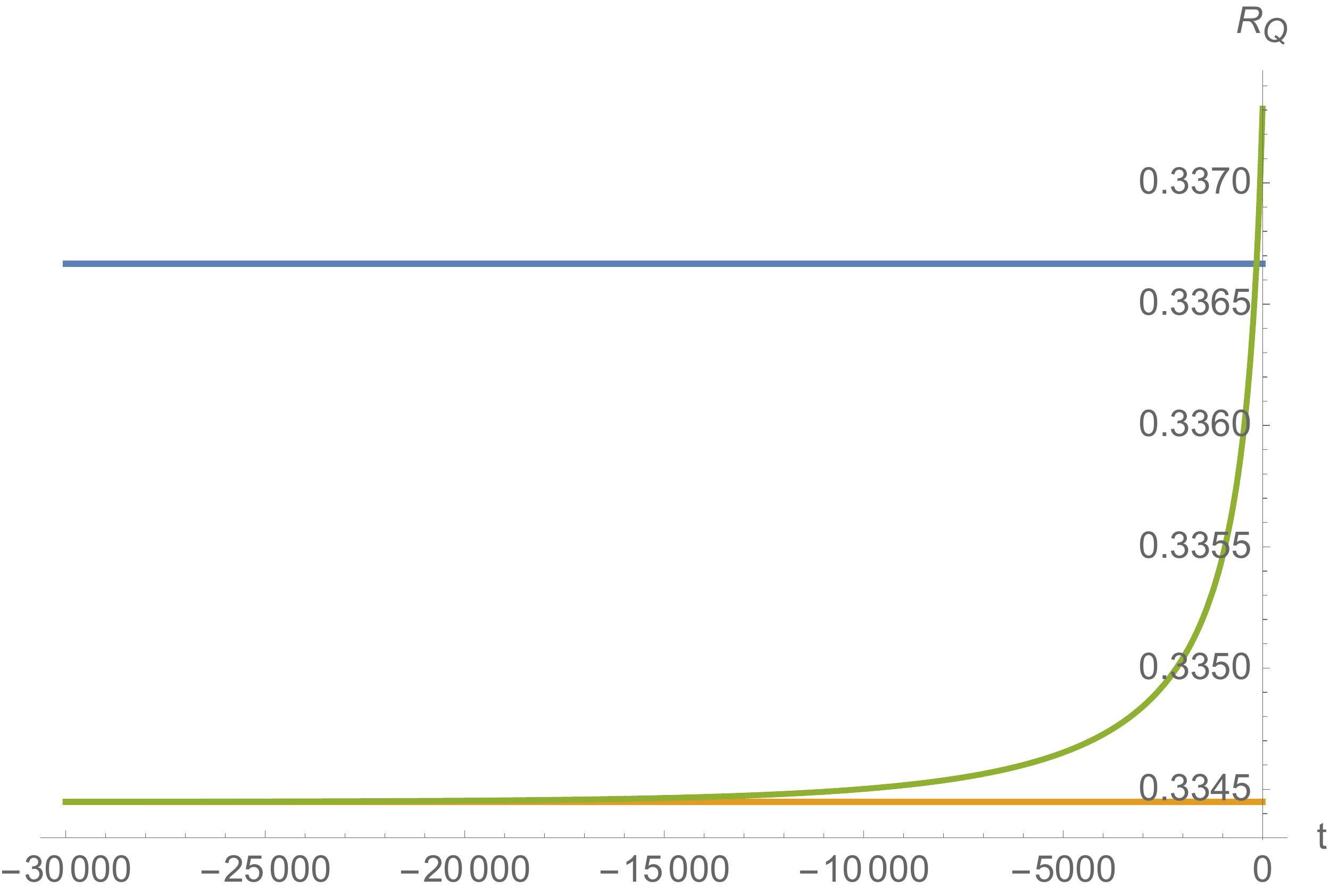} 
   \caption{\it 
  The R-charge (green) of the quark in the electric SQCD with gauge group SU($N_c$) 
  and $N_f$ quarks $Q+\tilde Q$, with $N_f=3\tilde N_c-1$, using (\ref{aEaM}). As before, the blue 
  straight line is the value at $t=0^+$, while the orange line is the asymptotic value in the IR.}
   \label{RQel}
\end{figure}

We will adopt the assumption, motivated in the Introduction, that in regions where the beta functions are small the $\aks$-function is the same as the function $\ak$ derived using the  
Lagrange multiplier definition \cite{Kutasov:2003ux}. Hence using \eqref{akfromR} and equating $\ak$'s in the two descriptions as in the Appendix, one finds 
\begin{align}
\label{aEaM}
2(N_c^2-1)~+~ 2N_fN_c \left(a_1(R_Q(t))-(R_Q(t)-R_Q(-\infty))a_1'(R_Q(t))\right) &\hspace{-1cm}\non\\
 ~~~~= ~~2(\tilde N_c^2-1)+  2N_f\tilde N_c  (a_1(R_q(t))- (R_q(t)-R_q(-\infty))&a_1'(R_q(t)) \non\\
~~~~~~+~N_f^2 (a_1(R_\Phi(t))-(R_\Phi(t)-R_\Phi(-\infty))a_1'(R_\Phi(t)))&~,
\end{align}
which can be used to determine $R_Q(t)$. Its behaviour is shown in Fig.~\ref{RQel}.

From there it is straightforward to determine the gauge coupling from the NSVZ beta function. Defining the electric gauge coupling as
\beq
\alpha_g~\equiv~\frac{N_cg^2}{(4\pi)^2}~,
\eeq
and using
\beq
\beta(\alpha_g)~=~-6\alpha_g^2f(\alpha_g)\left(1+\frac{N_f}{N_c}\left(R_Q-1\right)\right)~,
\eeq
one can now integrate, to find 
\beq
\label{DeltaF}
F(\alpha_g(t))-F(\alpha_g(0^-))~=~\int_0^tdt'\left(1+\frac{N_f}{N_c}\left(R_Q(t')-1\right)\right)~,
\eeq
where
\beq
F(x)\equiv\frac{1}{6}\left(\frac{1}{x}+2\log{x}\right)~.
\eeq
This can then be solved for $\alpha_g$. Note that as mentioned the $ {\cal O}(\Delta^2)$ terms in Eq.\eqref{Rqalpha} are required 
here. If they are omitted then there are order $1/N_c^2$ errors in the integrand, which over the order $-t\sim N_c^2$
 running required to get to the fixed point, translates into errors of order unity: in other words there would not be proper convergence to a fixed point. 
 
\begin{figure}[h] 
   \centering
   \includegraphics[width=4.in]{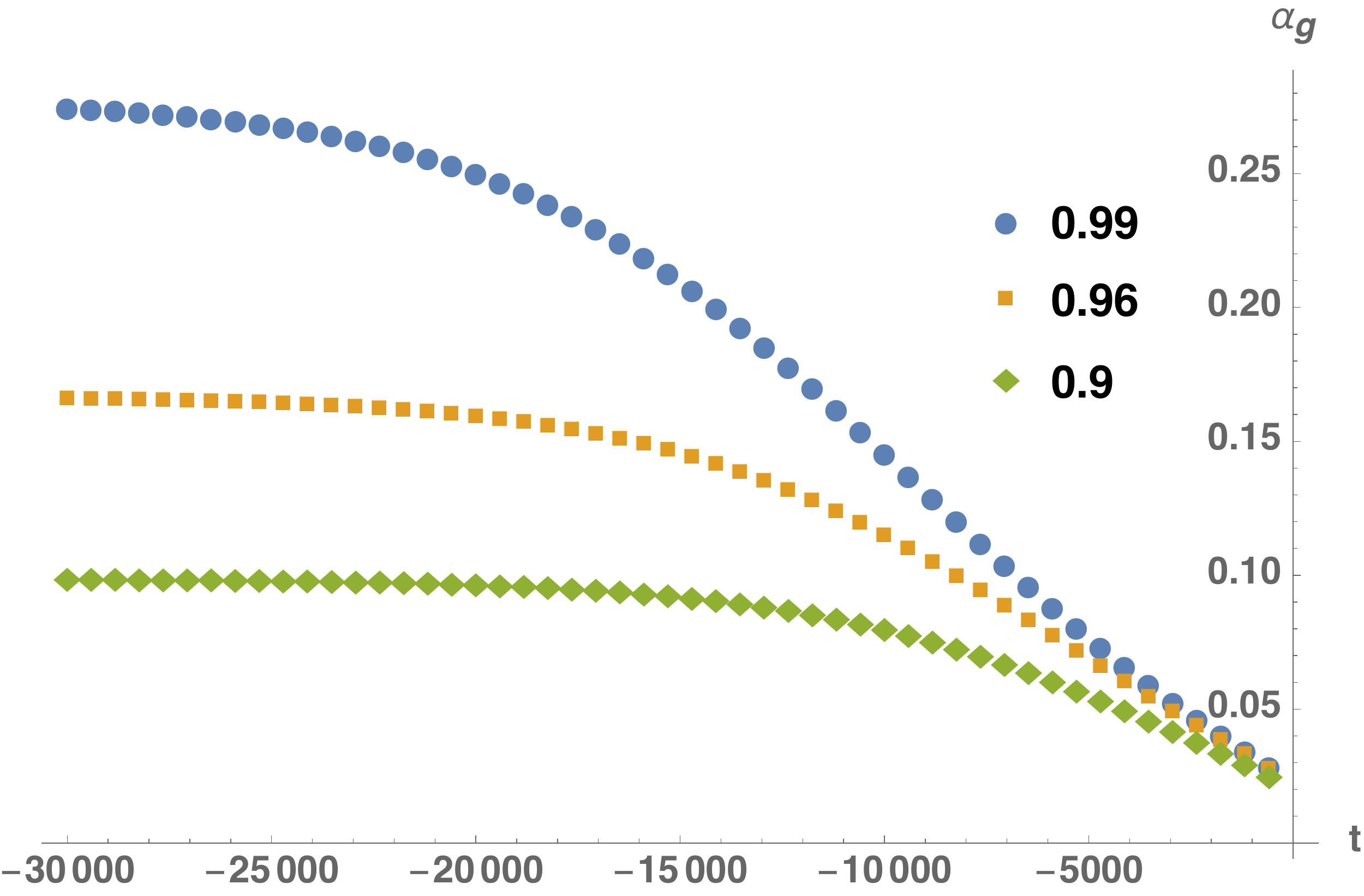} 
   \caption{\it The non-perturbative running of the gauge coupling constant of the 
   electric theory SU($N_c$) with $N_f$ quarks $Q+\tilde Q$, for three different values of 
   the ratio $\alpha_g(0^-)/\alpha_g^{max}(0^-)$, and for $\tilde N_c=100$, $N_f=3\tilde N_c-1$ 
   and $N_c=N_f-\tilde N_c$. Note that $\alpha_g~\equiv~\frac{N_cg^2}{(4\pi)^2}~.
$}
   \label{gel}
\end{figure}

Of course we do not know the numerical value of the boundary condition, $\alpha_g(0^-)$, in the electric theory, but since 
the r.h.s. of (\ref{DeltaF}) is negative, and since the gauge coupling must obey $\alpha_g<1/2$ 
in order that $f(\alpha_g)$ defined in (\ref{f}) does not change sign, there is a maximum allowed value of 
$\alpha_g(0^-)$ given by
\beq
F(\alpha_g^{max}(0^-))+\int_0^{-\infty}dt'\left(1+\frac{N_f}{N_c}\left(R_Q(t')-1\right)\right)~=~F(1/2)~.
\eeq
For our inputs this is given by
\beq
\alpha_g^{max}(0^-)~=~0.0216164~.
\eeq
As an illustrative example we take three different inputs $(0.99,0.96,0.9)$ for the ratio $\alpha_g(0^-)/\alpha_g^{max}(0^-)$ and
obtain numerically the  flows shown in Fig.~\ref{gel} for the non-perturbative coupling $\alpha_g(t)$. 
There is of course only one correct 
numerical boundary condition at $t=0^-$ corresponding to the electric theory dual to the 
perturbative magnetic one, but unfortunately it cannot be determined\footnote{It would be interesting to attempt to extend the approach to 
include the mass term explicitly with another Lagrange multiplier, as for the free-field theory in the introduction. However
as one would have to describe a Higgsing in the magnetic theory, this would be significantly more complicated, and it is not clear how 
one could fix several Lagrange multipliers with only a single $a$-parameter.}. All we know is that it must be non-perturbative, so 
too small ratios $\alpha_g(0^-)/\alpha_g^{max}(0^-)$ are unacceptable because they would not reproduce the known 
anomalous dimensions in the deep IR.

The entire flow including the $R$-charges can of course be expressed in terms of the Lagrange multipliers of \cite{Kutasov:2003ux}, in the manner described in the introduction and in Section~\ref{sec:ava}. We included them for completeness in the Appendix.   

\section{Equality of critical exponents}

The critical exponent provides a mild but nevertheless important check on the consistency of this picture. It is defined as the minimal eigenvalue of the matrix 
of coupling derivatives of the beta functions around the fixed point\footnote{Here one cannot compare the full matrices or 
even all their eigenvalues, because for example the dimensions of the matrices do not agree. However the minimal eigenvalue has a 
physical scheme-independent meaning in both descriptions.}
:
\beq
\beta^\prime\equiv{\rm min}\left\{{\rm positive\;eigenvalues}\left(\frac{\partial\beta_{\alpha_a}}
{\partial\alpha_b}\right)_{\rm F.P. }\right\}~.
\eeq
It is a renormalization scheme independent quantity and therefore should be equal for dual theories 
\cite{Anselmi:1996mq,Barnes:2004jj}. Usually of course this equivalence cannot 
be checked because one cannot compute in the strongly coupled theory. However our prescription 
($\aks^{\scriptscriptstyle (el)}=\aks^{\scriptscriptstyle  (mag)}$) allows it to be checked explicitly, as we now show.

In the magnetic description, the theory is perturbative and so we can simply use (\ref{rgeIRymag}) to evaluate the critical 
exponent:
\beq
\left(\frac{\partial\beta_{\alpha_a}}
{\partial\alpha_b}\right)_{\rm F.P. }~=~
\bem
-2\tilde\alpha_g^2(-\infty)\left(6-4\frac{N_f}{\tilde N_c}+2\frac{N_f}{\tilde N_c^3}\right) & 
-2\tilde\alpha_g^2(-\infty)\left(2\left(\frac{N_f}{\tilde N_c}\right)^2\right) \cr
2\tilde\alpha_y(-\infty)\left(1-\frac{1}{\tilde N_c^2}\right)(-2) & 2\tilde\alpha_y(-\infty)\left(2\frac{N_f}{\tilde N_c}+1\right)
\eem~.
\eeq
For $N_f=3\tilde N_c-1$ one obtains the leading order in $1/\tilde N_c$ approximation, 
using (\ref{alphayIR}):
\beq
\beta_{mag}^\prime~=~\frac{7}{3\tilde N_c^2}~,
\eeq
while the second, larger, eigenvalue is found to be equal to $14/(3\tilde N_c)$. 

In the strongly coupled electric description, there is a single gauge coupling, so that 
\beq
\label{betaprime}
\beta_{el}^\prime~=~\left.\frac{\partial\beta_{\alpha_g}(t)}{\partial\alpha_g(t)}\right|_{t\to-\infty}
~=~\left.\frac{\frac{d}{dt}\beta_{\alpha_g}(t)}{\frac{d}{dt}\alpha_g(t)}\right|_{t\to-\infty}
=~\left.\frac{d}{dt}\log\left(R_Q(t)-R_Q(-\infty)\right)\right|_{t\to-\infty}~.
\eeq
Usually in the non-perturbative theory the relation 
between $R_Q(t)$ and $\alpha_g(t)$ is not known. Here however we have a 
relation between $R_Q(t)$ and the known  $R_q(t)$ and $R_\Phi(t)$ 
of the perturbative magnetic theory through (\ref{aEaM}). We may therefore expand $a$ around $t=-\infty$,
\beq
a(t)~=~a(-\infty)+
%\sum_i\frac{\partial a}{\partial R_i}(R_k(-\infty))\left(R_i(t)-R_i(-\infty)\right)+
\sum_{i,j}\frac{1}{2}\frac{\partial^2 a}{\partial R_i\partial R_j}(-\infty)\left(R_i(t)-R_i(-\infty)\right)\left(R_j(t)-R_j(-\infty)\right)
+\ldots~,
\eeq

\noi
and from there must find
\bea
\frac{\partial^2a_{el}}{\partial R_Q^2}(-\infty)\left(R_Q(t)-R_Q(-\infty)\right)^2&~\approx~&
\frac{\partial^2a_{mag}}{\partial R_q^2}(-\infty)\left(R_q(t)-R_q(-\infty)\right)^2\non\\
&+&
\frac{\partial^2a_{mag}}{\partial R_M^2}(-\infty)\left(R_\Phi(t)-R_\Phi(-\infty)\right)^2~.
\eea
Since the second derivative of $a$ over the $R$-charges is proportional to the 
$b$-central charge of a conserved current (in this case it is the baryon current) 
and thus strictly non-zero, we must have the same scaling, 
\beq
R_Q(t)-R_Q(-\infty)~\sim~ R_q(t)-R_q(-\infty)~\sim~ R_\Phi(t)-R_\Phi(-\infty)~\sim~\exp{(\beta_{\scriptscriptstyle mag}^\prime t)}~,
\eeq
in the asymptotic region $t\to-\infty$. But then from (\ref{betaprime}) we consistently find 
\beq
\beta_{el}^\prime~=~\beta_{mag}^\prime~.
\eeq

We conclude that $a_{el}=a_{mag}$ along the flow is compatible with the equality of the electric and 
magnetic critical exponents. Of course this is not a particularly restrictive condition,  and many other relations would have given equality. For example  
\beq
a_{mag}~=~A(a_{el})
\eeq
for an arbitrary function $A(x)$ with  
\bea
%\label{A}
A(a_{el}(-\infty))&=&a_{el}(-\infty)~,\nonumber \\
\label{Aprime}
A^\prime(a_{el}(-\infty))&\ne&0~.
\eea
would suffice. 

\section{Conclusion}

In this paper we discussed the use of the $a$ central charge as a method of determining the flow in 
a strongly coupled supersymmetric theory from its weakly coupled dual. Although there are other examples of exact 
duality in field theory along an entire flow (e.g. \cite{Kapustin:1999ha}) this method seems particularly 
general and well suited to ${\cal N}=1$ supersymmetry.
Crucial to the approach is the equivalence of the scale-dependent $a$-parameter 
determined from the four-dilaton amplitude with an IR cut-off, and the $a$-parameter determined in the Lagrange multiplier method of Ref.\cite{Kutasov:2004xu,Kutasov:2003ux} with ``flowing'' $R$-charges.
We showed that this equivalence holds directly for massive free ${\cal N}=1$ superfields, as well as weakly coupled SQCD.  Assuming it to hold generally amounts to a particularly physical choice of RG scheme, in which the running $R$-charges are always determined precisely from the  four-dilaton amplitude. In this scheme, which is clearly well defined regardless of which formulation is being used, one can map the flow of a weakly coupled magnetic dual to the original strongly coupled electric theory. The specific system we considered  was the well-known pair of original SQCD Seiberg duals, with the magnetic description (with weak gauge and Yukawa coupling constants) running perturbatively from a fixed point in the UV to a different fixed point in the IR due to a mass-deformation, and the electric SQCD dual running between strongly coupled fixed points due to a meson-induced Higgsing. 

We should add that the mapping only seems to work straightforwardly in the direction of 
magnetic to electric, as in that case there is only one $R$-charge to determine (namely that of the electric quarks), and there is only one parameter (namely the $a$-parameter) with which to do it. Mapping in the converse direction may be possible in conjunction with $a$-maximisation \cite{Intriligator:2003jj}, but is less obvious.

\subsubsection*{Acknowledgments}
We are extremely grateful to Colin Poole for interesting discussions. BB acknowledges the financial support from the Slovenian Research Agency (research core funding No.~P1-0035). The work of FS is partially supported by the Danish National Research Foundation under the grant DNRF:90. BB thanks CP3 Origins Odense for hospitality.

\begin{appendices}
%\appendixpageoff

\section{The Lagrange multipliers}

Here we explicitly show how the Lagrange multipliers of \cite{Kutasov:2003ux,Kutasov:2004xu} flow in the model discussed in Section \ref{sec:flow}. 
We start with the original magnetic $a$-function,
\bea
\label{am}
\tilde a_{mag}&=&2\left(\tilde N_c^2-1\right)+2\tilde N_cN_fa_1(R_q)+N_f^2a_1(R_\Phi)\\
&&-~\tilde\lambda_gN_f\left(R_q-R_q(-\infty)\right)+\tilde\lambda_yN_f\left(2(R_q-R_q(-\infty))+
(R_\Phi-R_\Phi(-\infty))\right)\, .\non
\eea
By $a$-maximisation we have \cite{Intriligator:2003jj} 
\beq
\frac{\partial\tilde a_{mag}}{\partial R_q}~=~
\frac{\partial\tilde a_{mag}}{\partial R_\Phi}~=~0~,
\eeq
which gives the following for the Lagrange multipliers: 
\bea
\label{lambdagmag}
\tilde\lambda_g&=&2\tilde N_ca_1^\prime(R_q)-2N_fa_1^\prime(R_\Phi)~,\\
\label{lambdaymag}
\tilde\lambda_y&=&-N_fa_1^\prime(R_\Phi)~.
\eea
Similarly for the electric theory one gets 
\beq
\label{ae}
\tilde a_{el}~=~2\left(N_c^2-1\right)+2N_cN_fa_1(R_Q)-\lambda_gN_f\left(R_Q-R_Q(-\infty)\right)~,
\eeq
giving 
\beq
\label{lambdag}
\lambda_g=2 N_ca_1^\prime(R_Q)
\eeq
Plugging (\ref{lambdagmag}) and (\ref{lambdaymag}) into (\ref{am}), (\ref{lambdag}) 
into (\ref{ae}), and equating the two $a$-central charges, we obtain (\ref{aEaM}). 

\begin{figure}[h] 
   \centering
   \includegraphics[width=2.8in]{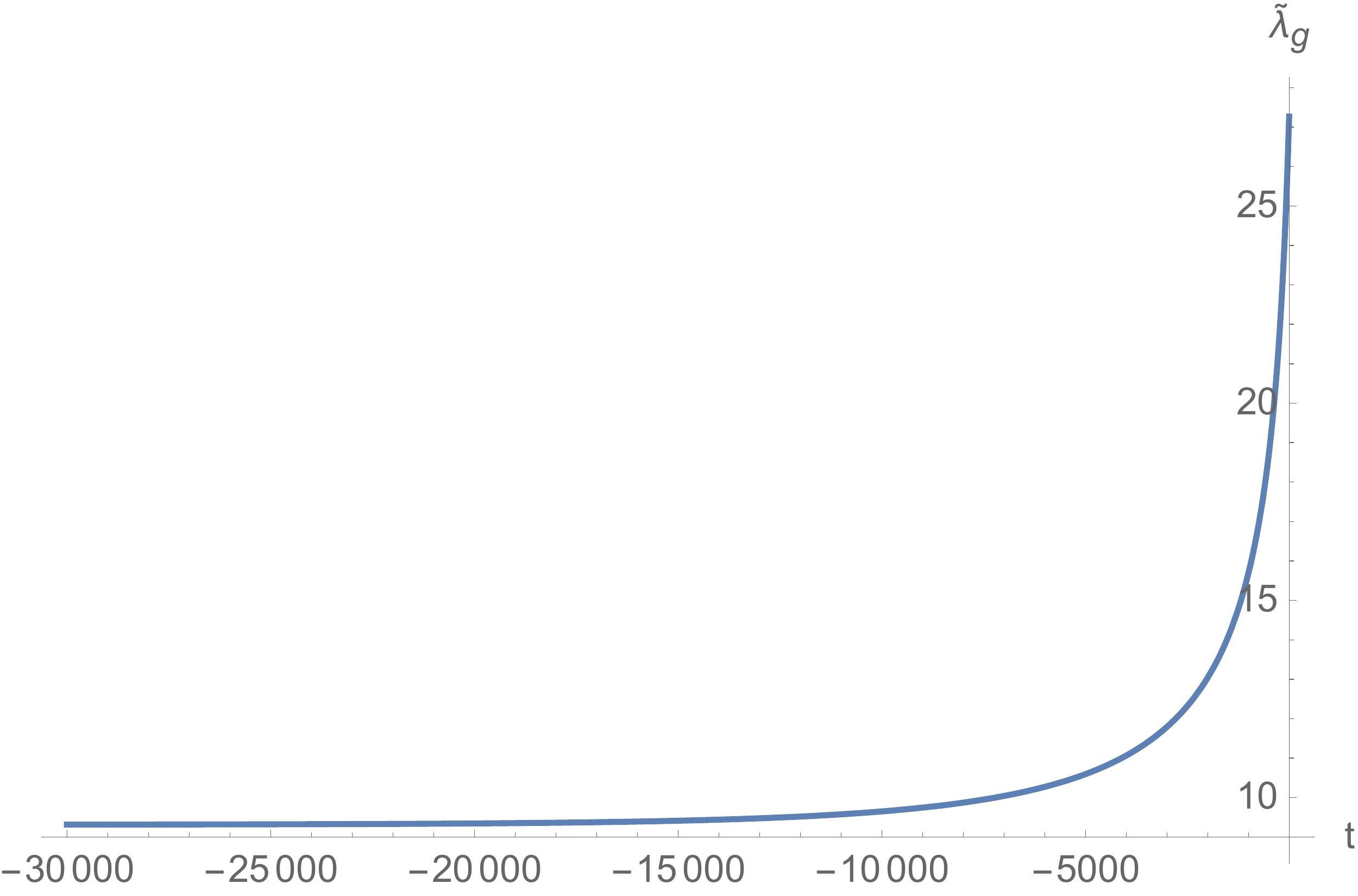} 
   \includegraphics[width=2.8in]{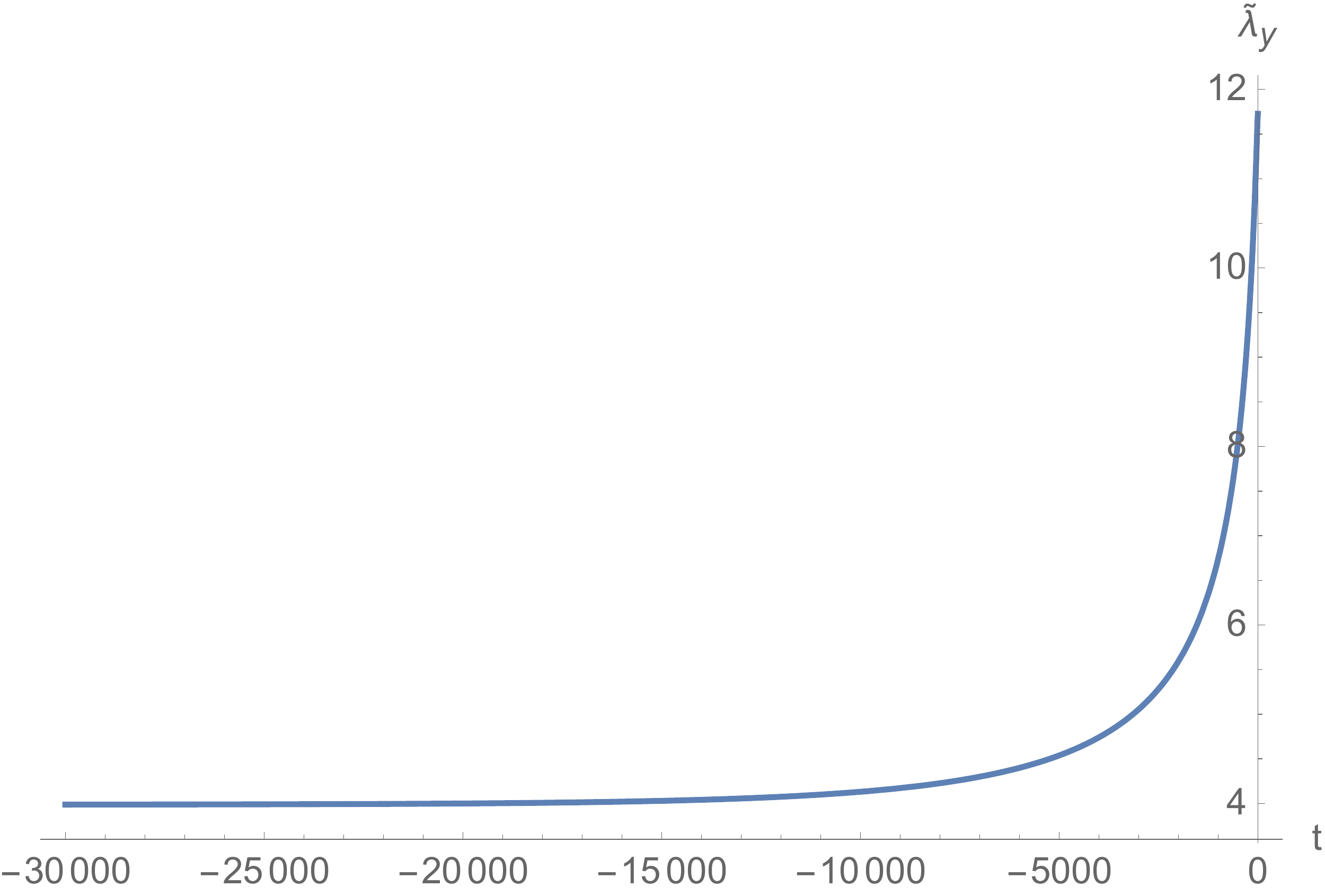} 
   \caption{\it 
The Lagrange multipliers of the magnetic theory.}
   \label{lglymag}
\end{figure}

\begin{figure}[h] 
   \centering
   \includegraphics[width=2.8in]{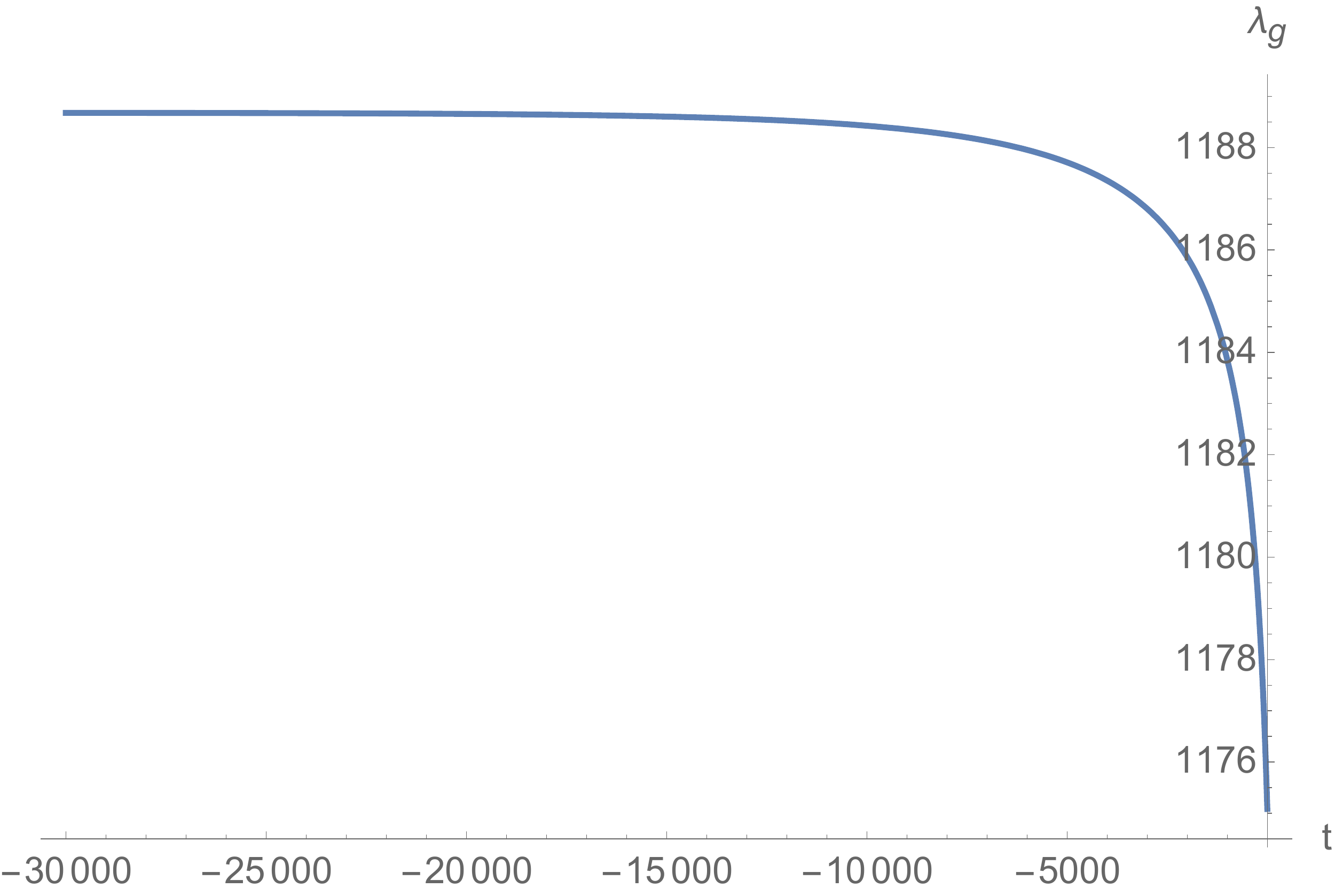} 
   \caption{\it 
The Lagrange multiplier of the electric theory.}
   \label{lgel}
\end{figure}

From the perturbative knowledge of $R_q(t)$ and $R_\Phi(t)$ we can thus draw 
$\tilde\lambda_g(t)$ in the magnetic theory discussed in the main text, while from the non-perturbative knowledge of 
$R_Q(t)$ using (\ref{aEaM}) we get $\lambda_g(t)$ for the strongly coupled electric theory. The graphs are shown in figs. \ref{lglymag} and \ref{lgel}.

\end{appendices}

\end{document}